\documentclass[12pt,aasms4]{article}

\usepackage{graphicx}
\usepackage{amsmath}   
\usepackage{amssymb}
\usepackage[comma,sort&compress]{natbib}

\def\eq#1{{Eq.~(\ref{#1})}}

\def\gsim{\;\rlap{\lower 2.5pt \hbox{$\sim$}}\raise 1.5pt\hbox{$>$}\;}
\def\lsim{\;\rlap{\lower 2.5pt  \hbox{$\sim$}}\raise 1.5pt\hbox{$<$}\;}

\begin{document}

\Large
\begin{center}
{\bf Signatures of Population III supernovae at Cosmic Dawn:  the case of GN-z11-flash} 
\end{center}

\medskip

\normalsize
\centerline{Hamsa Padmanabhan$^{1}$ \& Abraham Loeb$^{2}$}
\medskip

\centerline{\it $^{1}$ D\'epartement de Physique Th\'eorique, Universit\'e de Gen\`eve }
\centerline{\it 24 quai Ernest-Ansermet, CH 1211 Gen\`eve 4, Switzerland}
\centerline{email: hamsa.padmanabhan@unige.ch}

\centerline{\it $^{2}$ Astronomy department, Harvard University}
\centerline{\it 60 Garden Street, Cambridge, MA 02138, USA}
\centerline{email: aloeb@cfa.harvard.edu}

\vskip 0.2in
\hrule
\vskip 0.2in
  
\begin{abstract}
We illustrate the observability of the end stages of the earliest (Population III) stars at high redshifts $z \gtrsim 10$, using the recently observed transient, GN-z11-flash as an example. We find that the observed spectrum of this transient is consistent with its originating from a shock-breakout in a Population III supernova occurring in the GN-z11 galaxy at $z \sim 11$. The energetics of the explosion indicate a progenitor star of mass $\sim 300 M_{\odot}$  in that galaxy, with of order unity such events expected over an observing timescale of a few years. We forecast the expected number of such transients from $z > 10$ galaxies as a function of their host stellar mass and star formation rate. Our findings are important in the context of future searches to detect and identify the signatures of galaxies at Cosmic Dawn.
\end{abstract}

 \section{Introduction}
The first stars in the Universe are believed to be massive, metal-free Population III (Pop III) stars \citep{abel2002, bromm2002}, which synthesized heavy elements (such as C, O, Ne, Mg, Si, and Fe) through their supernova explosions and  played a major role in the chemical evolution of the early Universe. We describe the observability of such explosions from very high redshifts in future searches such as with the \textit{James Webb Space Telescope} (JWST), using the recently published rest-frame ultraviolet flash reported by \citet{jiang2020} as a example.

GN-z11-flash --- a rest-frame ultraviolet flash with an observed duration of a few hundred seconds ---  was recently reported \citep{jiang2020} from GN-z11, a luminous, star forming galaxy situated at $z \sim 11$. We examine the scenario in which this transient originated from the shock breakout  taking place in a supernova explosion during the end stages of  a 300 $M_{\odot}$ Population  III (Pop III) star in the galaxy.\footnote{The alternative explanation of a gamma-ray burst afterglow for GN-z11-flash {{is}} much less likely given the much lower rate of occurrence for such events \citep{jiang2020, kann2020}.} The observed star formation rate \citep{jiang2020a, oesch2016} of the GN-z11 galaxy ($\sim 26 \ M_{\odot}$/yr) with a top heavy IMF (characteristic of the first stars) is consistent with $\sim 1$ Pop III supernova observed every few years.
 We forecast the expected number of such transients expected from future surveys based on extrapolating the empirically constrained star formation rates and stellar mass functions at lower redshifts. Our findings serve as a pointer towards informing the searches and follow-up of such transients by the JWST and similar missions.

 \section{Explosion mechanism}
 \label{sec:scenario}

 GN-z11 is a luminous galaxy at $z \sim 11$ with an intrinsic star formation rate of $26 \ M_{\odot}$/yr. The breakout of a shock \citep[e.g.,][]{chevalier1976, colgate1974} from a  supernova associated with a Population III star in GN-z11 would be detectable in the UV to {X-ray} bands. 
We can estimate the energy released in the shock breakout by  connecting the observed luminosity of the flare \citep{jiang2020}, the time of the shock breakout and the radius of the star. We begin with the connection between the peak observed luminosity of the flare, $L_{\rm obs}$, the radius of the star, $R_*$ and the internal energy of the breakout shell, $E_0$, which is given by:
 \begin{equation}
 \frac{E_0 c}{R_*} \equiv L_{\rm obs}= 1.3 \times 10^{47} \text{ergs/s}
 \label{obslum}
 \end{equation} 
 The timescale of the observed flare (defined by the time of transition between the planar and spherical geometries of the breakout shell) is given by $ t_s = R_*/v_0 \sim  35 \ \text{s}$,
 where $v_0$ is the velocity of the shock when it breaks out.\footnote{We have used the observed time, $t_{\rm s, obs} = 424 \ \text{s}$ scaled down by the redshift dilation factor $(1+z) \sim 12$. } We can relate $v_0$ to $E_0$ by invoking the energy released when the star collapses to form a black hole. This energy is delivered to the mass of the stellar envelope, $m_0$, through the radiative shock which heats it up \citep[e.g.,][]{nakar2010}. When the shock reaches the surface of the star, it heats the surface layer as well. The observer sees the flash only from the surface layer at an optical depth $\tau=c/v_0$ and all the radiation locked interior to that takes a long time to diffuse out. The flash represents the energy that can immediately leak out during the flash, defined by $E_0 \sim m_0 v_0^2$, where $m_0$ and $v_0$ are the mass  and velocity of the breakout shell respectively. 
 
 Population III stars are held against their self-gravity by radiation pressure, which maintains their radiation field at the Eddington luminosity. For a surface temperature of $T_{\rm eff} \sim 10^5$ K, their radius $R_*$ and mass $m_*$ can be connected by:
\begin{equation}
R_* \approx 4.3 \times 10^{11} \text{cm} \times \left(\frac{m_*}{100 M_{\odot}}\right)^{1/2}
\label{rstarmstar}
\end{equation}
{Given that the shock breakout timescale is of the order of 35 s in the rest frame, the shock needs to cross a region of radius at most 15 $R_{\odot}$.  We consider a  $\sim 10 R_{\odot}$ radius star, which leads to 
  $m_* \sim 300 \ M_{\odot}$ from \eq{rstarmstar}. The mass value of $300 M_{\odot}$, for the above surface temperature of $T_{\rm eff} \sim 10^5 {\rm K}$, corresponds to a stellar luminosity of $10^7 L_{\odot}$ \citep{schaerer2002}. }  The above parameters are consistent with those expected for post-main sequence Population III stars of a few hundred solar masses that undergo core-collapse \citep{ohkubo2009}.
  
 Using $R_* = 10 R_{\odot}$ in \eq{obslum}, we find $E_0 \sim 3 \times 10^{48}$ ergs. We then have $v_0 \sim R_*/35$ s $\sim 0.7 c$, which then leads to  the mass of the shell being given by:
\begin{equation}  
 m_0 = E_0/v_0^2 = 7 \times 10^{27} \text{g} \sim 10^{-8} m_*.
 \label{m0toE0}
\end{equation}

Given a power-law behaviour of the  stellar density profile with the radial distance, $\rho \sim (R_* - r)^{n}$, we can relate the mass of the shell and the stellar parameters by generalizing the treatment in \citet{nakar2010} {{to the case of an arbitrary polytropic index. We begin by noting that the density profile near the star surface can be approximated as:
\begin{equation}
\rho = \rho_*\left(\frac{d_i}{R_*}\right)
\end{equation}
where $d_i$ is the width of the shock breakout shell at the time of breakout, and $\rho_*$ is mean density, $\rho_* = m_*/R_*^3$. This is then used to derive the following properties of the breakout shell, namely the mass $m_0$, optical depth $\tau$ and velocity of the shell $v_0$, which are given by:
\begin{eqnarray}
m_0 &=& \frac{4 \pi R_*^3 \rho_*}{n + 1} \left(\frac{\rho}{\rho_*}\right)^{\frac{n+1}{n}} \nonumber \\
\tau_0 &=& \frac{\kappa R_* \rho_*}{n + 1} \left(\frac{\rho}{\rho_*}\right)^{\frac{n+1}{n}} \nonumber \\
v_0 &=& 1800 {\rm km/s} \left(\frac{E_{51}}{m_{15}}\right)^{1/2} \left(\frac{\rho}{\rho_*}\right)^{-0.19}
\end{eqnarray}
in terms of the density $\rho$ of the front shell of gas at the breakout time. In the above expressions, the opacity $\kappa = 0.2 (1+X)$ where $X = 0.76$ is the mass fraction of hydrogen \citep{bromm2001}. In the above expression, $m_{15} = (m_*/15 \ M_{\odot})$ and $E_{51} = (E/10^{51} \ \text{ergs})$, where $E$ is the total energy of the explosion. 
Setting $\tau_0 = c/v_0$, as required for the breakout, we obtain the general form of the relation between the mass of the breakout shell and the remaining parameters:

\begin{equation}
m_0 = \frac{4 \pi m_*}{(n + 1)} \left[A \left(\frac{m_{15}}{E_{51}}\right)^{1/2} \frac{(n+1) R_*^2}{\kappa \ m_*}\right]^{\beta (n+1)/n}
\label{m0equation}
\end{equation} 
 where $\beta = 1/((n+1)/n - 0.19), A = c/(1800  \ \text{km s}^{-1}) = 166.7$. The exponent $n$ is related \citep{calzavara2004} to the polytropic index $\gamma_p$ for the star as $\gamma_p = 1 + (1/n)$. For  a Pop III star  with polytropic index  $\gamma_p = 3$, this gives $n = 0.5$. 
 
Substituting the values of $m_*$ and $R_*$ for our present case into \eq{m0equation}, we can thus relate the value of $m_0$ to the total energy of the explosion, as:
\begin{equation}
m_{0} \sim 1.3\times 10^{-6}  E_{51}^{-0.54} M_{\odot}
\label{m0toE51}
\end{equation}
Finally, substituting for $m_0$ from \eq{m0toE0}, we obtain:
\begin{equation}
E_{51} \sim 0.3 
\end{equation}
implying an explosion energy of $E \sim 3 \times 10^{50}$ \rm {ergs}.}}  {The binding energy of a polytrope with $n = 0.5$ and the mass and radius inferred for the Pop III star is given by:
\begin{equation}
\Omega = 3 Gm_* ^2/(5 - n) R_* = 2.3 \times 10^{52} \rm{ergs}
\end{equation}
and can be easily supplied by the radiative efficiency of converting the accretion into the central black hole to photons and neutrinos.
}

 \section{Event rate}
 \label{sec:eventrate}

\begin{figure*}
\begin{center}
\hskip-0.5in \includegraphics[width = 0.5\textwidth]{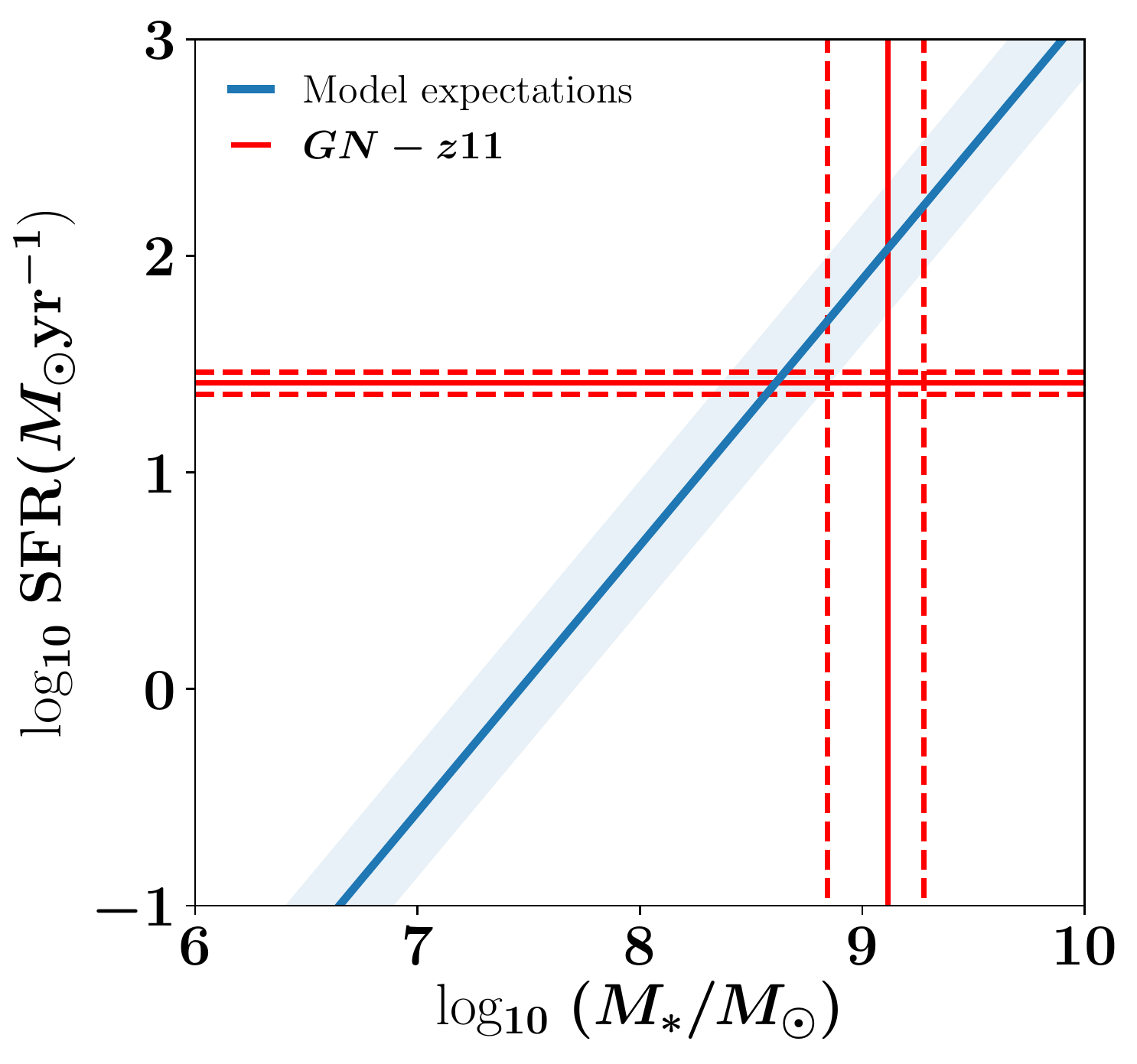}  \includegraphics[width = 0.5\textwidth]{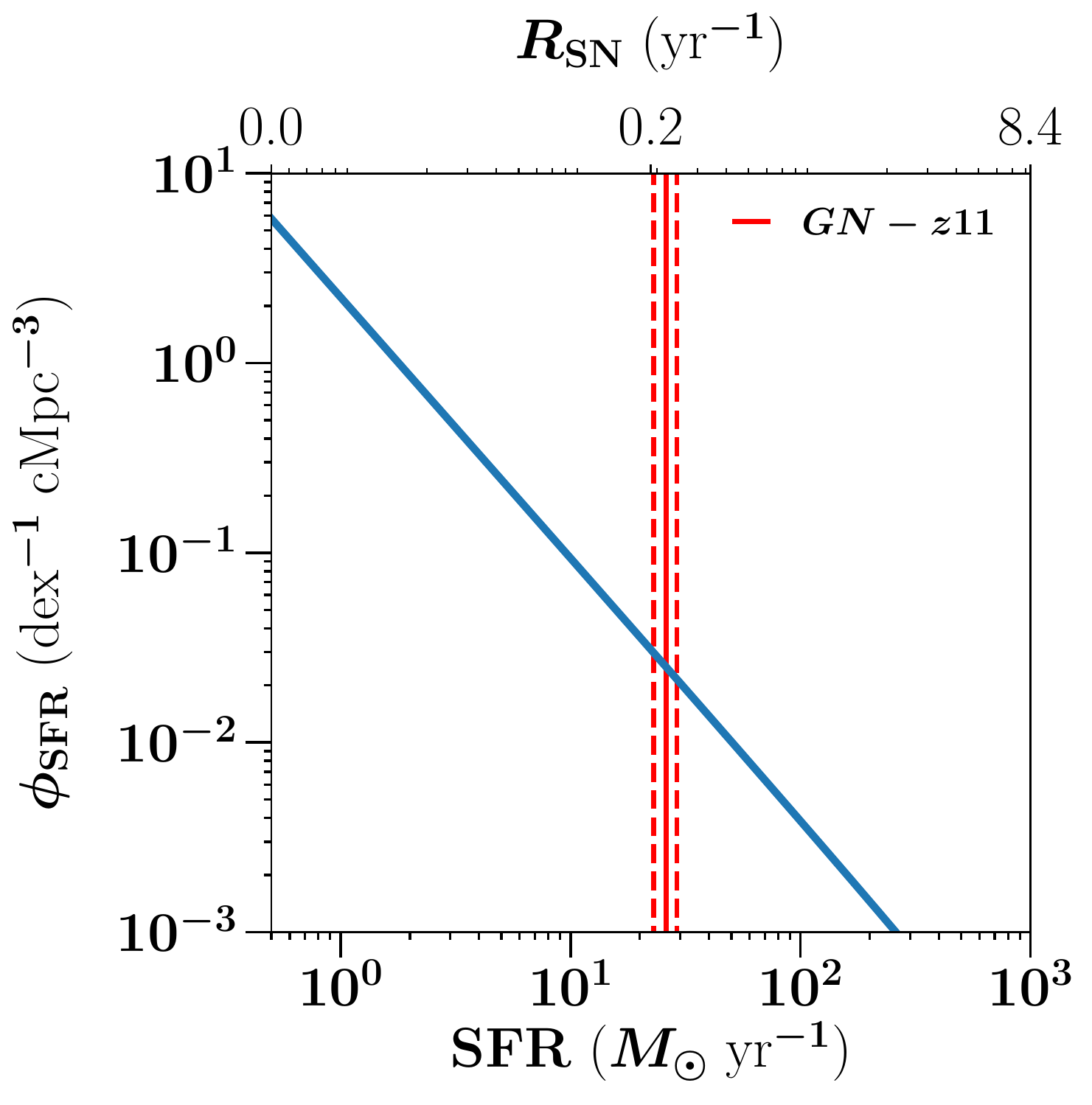}
\caption{\textit{Left panel:} Stellar mass to star formation rate relation for galaxies at $z \gtrsim 10$ (blue solid curve) and its associated uncertainty based on current observations \citep{behroozi2019, matthee2019} extrapolated to the observed stellar mass range of GN-z11. The inferred stellar mass and star formation rate of GN-z11 are shown by the red solid lines (with their uncertainties indicated by the dashed lines). \textit{Right panel:} Comoving number density of galaxies per dex of SFR obtained by combining the stellar mass function \citep{davidzon2017}, assuming negligible evolution, with the stellar mass - SFR relation in the left panel. The top $x$-axis shows the corresponding (observed)  number of  supernovae per year expected for a Pop III IMF at $z \sim 11$.}
\label{fig:sfrsnrate}
\end{center}
\end{figure*}
{{ In this section, we estimate the detectability of flares from Pop III explosions at $z \sim 11$ in galaxies similar to GN-z11.
To begin with, we can estimate the rate of  supernovae in such galaxies by using the relation from lower redshift observations calibrated in \citet{melinder2012}:
 \begin{equation}
 R = k  \ \text{SFR}(z)
 \label{supernovarate}
 \end{equation}
 where $R$ is the intrinsic supernova (SN) rate for a galaxy with star-formation rate SFR$(z)$ at redshift $z$, and the constant $k$ is the percentage of stars that explode as supernovae per unit mass. At low redshifts, it is found that $k = 0.014/M_{\odot}$, a typical value for the flattest initial mass function (IMF) as adopted by \citet{hopkins2006}. However, for Pop III stars at early times,  the findings of \citet{pan2012} predict a   core collapse SN rate of $10^{-3} \text{yr}^{-1} \text{Mpc}^{-3}$ for Pop III stars, for a star formation rate of 0.01 $M_{\odot} \text{Mpc}^{-3} \text{yr}^{-1}$. This leads to a proportionality constant of $k_{\rm Pop III} \sim 0.1/M_{\odot}$, nearly an order of magnitude higher than for galaxies in  the lower-redshift universe. 
 
For the case of GN-z11, whose star formation rate is ${\rm SFR} \sim 26 \ M_{\odot} \ {\rm yr}^{-1}$, \eq{supernovarate} predicts a near-unity detection of the supernova rate when observations are made over a period of a few years. A similar proportionality factor is obtained from the model of \citet{mebane2018}, with the SN rate of about $10^{-6} \text{yr}^{-1} \text{Mpc}^{-3}$ for a Population III IMF, with the corresponding  SFR being $10^{-5} M_{\odot}  \text{Mpc}^{-3} \text{yr}^{-1}$, thus leading to $k_{\rm Pop III} \sim 0.1/M_{\odot}$, resulting in of order unity supernovae detected over a few years' observation time.

 We can estimate the likelihood of observing such flares as a function of their star formation rate and stellar mass within a fixed comoving volume at $z \sim 11$. To do this, we use the empirically derived stellar mass - star formation rate connection \citep{behroozi2019} for $z \gtrsim 10$ galaxies, and combine this with the comoving number density of such galaxies expected at these epochs \citep{davidzon2017} assuming negligible evolution of the stellar mass function from $z \sim 5-10$. The results are shown in Figure \ref{fig:sfrsnrate}, with the inferred stellar mass [$(1.3 \pm 0.6) \times 10^9 M_{\odot}$] and star formation rate ($26 \pm 3 \ M_{\odot}$ yr$^{-1}$) of GN-z11 overplotted as the red lines. The top $x$-axis in the right panel also shows the corresponding observed supernova rate per year expected for a Pop III IMF, as discussed above, consistent with the estimate of $\sim 1$ detection of such an event within an observing timescale of a few years. 
 }}

 \section{Discussion}
In this paper, we have explored the observability of transient signatures from the end stages of massive Pop III stars at high redshifts, using the recently reported GN-z11-flash as an example. We examined the scenario in which the observed flare --- from the GN-z11 galaxy at $z \sim 11$ --- arises as a result of the shock breakout from a supernova explosion in a Pop III star located in this galaxy. A top heavy IMF predicts enough supernova events to be observed given the observed SFR of this and similar early galaxies.
 The simplest model is that of a Pop III star with mass $\sim 300 M_{\odot}$ and radius $\sim 10 R_{\odot}$, which core collapses to make a black hole, releasing a total energy $E \sim 3 \times 10^{50}$ ergs from the accretion of the core mass into the black hole. The part of the energy associated with the flash at the surface layer (with mass $m_0 \sim 10^{-6} M_{\odot}$) leads to the observed luminosity of $1.3 \times 10^{47}$ ergs/s. This event may thus present the first observational evidence for the death of very massive stars at $z \gtrsim 10$. 
 
 {We note that the mechanism leading to energy release in this case is not restricted to being a  supernova explosion, since the only relevant parameter involved is the total energy of the explosion. If the 300 $M_{\odot}$ star is prevented from collapsing into a black hole with an explosion, as may be suggested by some simulations, the arguments of Sec.\ref{sec:scenario} would continue to hold, since the only required observables constraining the mechanism are the luminosity and time duration of the flare, as related to the host's stellar mass and radial profile. It is not possible to obtain the required energetics with a smaller mass star or a different central engine such as, e.g., a magnetar-powered explosion, as we describe in the Appendix. Other possible scenarios for the energy release include choked gamma-ray bursts (GRBs) \citep[e.g.,][]{piran2017, izzo2019} with the explosion powered by the black hole following, e.g., a Type 1bc supernova explosion, {or a Type 1c explosion \citep[e.g.,][]{shankar2021}}. Choked GRBs may be much more numerous than successful GRBs, especially at high redshifts, thus leading to a much larger event rate \citep[e.g.,][]{razzaque2003, meszaros2001, denton2018, ando2005}. {A superluminous supernova (SLSN)-like explosion taking place in a Population III star is also in line with the low metallicities expected \citep[e.g.,][]{galyam2009, angus2016} and energy of the explosion \citep[e.g.,][]{nicholl2017} found for GN-z11-flash. Hydrogen-poor SLSNe hosts \citep{leloudas2015} have been found to have host galaxy stellar masses and specific star formation rates (sSFRs) consistent with the values observed in GN-z11.}
}
 
 The progenitor stars of these flares (shock breakout events) exist mainly at $z>10$ in low-metallicity galaxies. Flares such as GN-z11-flash from supernovae at Cosmic Dawn  \citep[e.g.,][]{moriya2010, kasen2011, tanaka2013, desouza2013, loeb2013, whalen2013a, whalen2013b} could make galaxies that are otherwise too faint, detectable, leading to a new method to flag $z>10$ galaxies by monitoring flares from them. For example, it is known that unlensed Pop III stars  at $z \sim 7-17$ are not directly observable with the \textit{James Webb Space Telescope} \citep{windhorst2018}, with lensed ones also being extremely difficult to detect \citep{rydberg2013}. Searching for  flaring $z>10$ galaxies by taking multiple snapshots of the same region of the sky will be important to detect and identify similar signatures of the end stages of Pop III stars, and would lead to a detailed understanding of the properties of their progenitors.

\section*{Acknowledgements} HP acknowledges support from the Swiss National Science Foundation under the Ambizione Grant PZ00P2\_179934. The work of AL was partially supported by Harvard's Black Hole Initiative, which is funded by grants from JTF and GBMF.

\section*{Data availability} The manuscript has no associated data.

\section*{Appendix}
In the main text, we have considered the shock breakout associated with a core-collapse supernova from a massive Pop III star, whose energetics yield an explosion energy consistent with expectations.
For completeness, we provide below the analyses for  (i) a smaller mass star than considered in the main text, and (ii) a possible magnetar powered explosion. We find that both these cases have a lower likelihood of powering the GN-z11-flash compared to the 300 $M_{\odot}$ case considered in the main text.

(i) \textit{ A smaller mass  supernova explosion}. For the case of a $\sim  100 \ M_{\odot}$ star which collapses to form a black hole, the same arguments as described in the main text yield the following values for the parameters of the explosion:

\begin{eqnarray}
m_* &\sim & 100 M_{\odot} \nonumber \\
R_* &\sim &  6 R_{\odot} \nonumber \\
E_0 &\sim & 1.9 \times 10^{48} \rm{ergs} \nonumber \\
v_0 &\sim & 0.4c \nonumber \\
m_0 &\sim & 1.3 \times 10^{28} \ g \sim 6.6 \times 10^{-8} m_*  
\end{eqnarray}
Using \eq{m0toE51} of the main text, we find:
\begin{equation}
E_{51} \sim 0.003;
\end{equation}
implying that the energy release associated with the supernova is:
\begin{equation}
E \sim 3 \times 10^{48} \  \rm{ergs}  ,
\end{equation}
which, in turn, suggests that about half the energy release goes into the shell at the shock outburst. Thus, supernova  explosions with the above mass range do not provide a likely scenario for the observed flare. Explosions in stars of smaller masses are likely to drive the energy release to lower values, making them less probable compared to the 300 $M_{\odot}$ case considered in the main text.

(ii) \textit{A possible magnetar-powered explosion. }
Assuming that the luminosity of the GN-z11-flash corresponds to the peak luminosity of the breakout, we obtain a rotational energy \citep[][]{kasen2010}:

\begin{equation}
E_p \sim 2 \times 10^{52} P_{1ms}^{-2} \ \rm{ergs} \,
\end{equation}
in terms of the period $P_{1ms}$ of the pulsar in units of 1 ms, 
and the characteristic time \citep[e.g.,][]{kasen2016}
\begin{equation}
t_p \sim 5 B_{14}^{-2} P_{1 ms}^2 \  \rm{days} \, ,
\end{equation}
where $B_{14}$ is the magnetic field in units of $10^{14} \rm{G}$.
This means that, even with a smallest possible period of 1 ms for the pulsar, $t_p \sim 35$ sec  for the flash requires the magnetic field to be
$B \sim 100 \times 10^{14} \rm{G} \sim 10^{16} \rm{G}$.
Although this is just within the limit for a strong magnetar, we also need $E_p \sim  E_{\rm rad} \sim  10^{52} \rm{ergs}$, which corresponds to the total radiated energy. In that case \citep{kasen2010}, the resulting peak luminosity is:
\begin{equation}
L_{\rm peak} \sim 5 \times 10^{43} B_{14}^{-2} E_{51}^{1/2} \rm{ergs/s}
\end{equation}
assuming the ejected mass $M_{\rm ej} \sim 5 M_{\odot}$ and the opacity $\kappa \sim 0.1$.
However, for the above value of $B \sim 10^{16} \rm{G}$, this leads to an unnaturally large $E_{51}$, i.e. requiring $L_{\rm peak} \sim 10^{47} \rm{ergs/s}$ leads to $E_{51} \sim 10^{15}$, which is too high.
This is also the case following Eqs. 26-27 of \citet{kasen2016},  since requiring the shock breakout time, $t_{\rm bo} \sim 35$ sec forces an unnaturally high $E_{51}$ for a peak luminosity of $10^{47}$ ergs, assuming other parameters are set to their default values.
Hence, we conclude that a magnetar-powered scenario is also unlikely to explain the observed flare.

\def\aj{AJ}                   
\def\araa{ARA\&A}     
\def\eprint{}        
\def\apj{ApJ}                 
\def\apjl{ApJ}                
\def\apjs{ApJS}               
\def\ao{Appl.Optics}          
\def\apss{Ap\&SS}             
\def\aap{A\&A}                
\def\aapr{A\&A~Rev.}          
\def\aaps{A\&AS}              
\def\azh{AZh}                 
\def\baas{BAAS}
\def\jcap{JCAP}
\def\jrasc{JRASC}             
\def\memras{MmRAS}
\def\na{New Astronomy}
\def\nat{Nature}
\def\mnras{MNRAS}             
\def\pra{Phys.Rev.A}          
\def\prb{Phys.Rev.B}          
\def\prc{Phys.Rev.C}          
\def\prd{Phys.Rev.D}          
\def\prl{Phys.Rev.Lett}       
\def\pasp{PASP}               
\def\pasj{PASJ}
\def\physrep{Phys. Repts.}
\def\qjras{QJRAS}             
\def\skytel{S\&T}             
\def\solphys{Solar~Phys.}     
\def\sovast{Soviet~Ast.}      
\def\ssr{Space~Sci.Rev.}      
\def\zap{ZAp}                 
\let\astap=\aap
\let\apjlett=\apjl
\let\apjsupp=\apjs

\small{
\bibliographystyle{aa}
\bibliography{mybib}
}

 \end{document}